# Strain-tunable multipiezo effects in Janus monolayer Cr$_2$SSe: Selective reversal of valley polarization and single-spin-channel anomalous valley Hall effect

Quan Shen, Jianing Tan, Tao Yao, Wenhu Liao, and Jiansheng Dong*

*Department of Physics, Jishou University, Jishou 416000, Hunan, China*
*Corresponding authors: jsdong@jsu.edu.cn

## ABSTRACT

Altermagnetism, the third class of collinear magnetic order, uniquely combines a zero net magnetization with spin-polarized bands in reciprocal space, opening new avenues for two-dimensional valleytronics and spintronics. Here, using first-principles calculations, we predict that the Janus monolayer Cr$_2$SSe, which possesses intrinsic inversion symmetry breaking, hosts a strain-tunable multipiezo effect and exhibits distinctive valleytronic properties. The system displays pronounced spin splitting and band inversion at the X and Y high-symmetry points in the Brillouin zone, giving rise to robust spin-valley locking. The degeneracy of these valleys is protected by diagonal mirror symmetry. Application of uniaxial strain breaks this symmetry, concurrently inducing piezovalley, piezoelectric, and piezomagnetic responses, a manifestation of the multipiezo effect. Critically, strain applied along orthogonal crystallographic directions yields opposite valley polarization, while under small compressive strain, we achieve selective reversal of valley polarization, enabling independent control of valence and conduction band valleys and promoting a single-spin-channel anomalous valley Hall effect. These findings establish a pathway for low-power, non-volatile manipulation of valley degrees of freedom and enhanced spin transport efficiency, providing a theoretical foundation for the design of energy-efficient valleytronic devices.

# I. INTRODUCTION

The pursuit of electronic architectures beyond conventional limits has driven intensive research into magnetic materials that enable precise control over spin-dependent phenomena [1-3]. While two-dimensional antiferromagnets (AFMs) offer inherent advantages for spintronic applications, such as ultrafast dynamics and immunity to stray fields, their spin-degenerate band structure presents a fundamental obstacle to generating and manipulating spin-polarized currents [4,5]. This landscape has been transformed by the discovery of altermagnetism (AM), a distinct class of collinear magnetic order. AMs achieve a compensated ground state with zero net magnetization, yet exhibit alternating spin polarization in reciprocal space through the breaking of combined parity-time (*PT*) symmetry [6,7]. By uniquely integrating characteristics of ferromagnetic (FM) and AFM systems, they unlock distinctive properties, including spin-dependent transport, non-zero Berry curvature, and valley-contrasting physics, establishing them as compelling platforms for high-performance quantum devices [8,9].

Two-dimensional magnetic materials, by virtue of their atomic-scale thickness and exceptional mechanical flexibility, provide an ideal platform for strain engineering. These attributes enable precise modulation of multiple coupled degrees of freedom, including magnetic, electronic, and valley properties, offering extensive opportunities for investigating emergent piezo-induced phenomena [10-13]. Current research in this domain focuses primarily on three classes of piezo-induced effects: piezovalley, piezoelectric, and piezomagnetic responses. First discovered by Jacques and Pierre Curie in 1880, the piezoelectric effect establishes a bidirectional coupling between mechanical and electrical degrees of freedom: piezoelectric materials develop an electric polarization in response to applied strain and, conversely, undergo mechanical deformation when subjected to an applied voltage [14,15]. In a related manner, piezomagnetic materials, including those with zero net spontaneous magnetization, exhibit a modulation of their magnetic moment in response to mechanical stress, or, conversely, display magnetostrictive deformation under an external magnetic field [16,17]. By contrast, the piezovalley effect refers to the generation of valley polarization under mechanical strain via the lifting of valley degeneracy, a mechanism of central importance to valleytronics [18-20].

Recently, several piezo-induced phenomena have been reported in two-dimensional AM systems. For instance, the piezoelectric effect can be realized in these materials through the construction of Janus structures, whose intrinsic inversion symmetry breaking gives rise to pronounced electromechanical coupling [21,22]. In contrast, the piezomagnetic effect is comparatively rare in two-dimensional AMs and typically requires the combined influence of uniaxial strain and carrier doping. This interplay disrupts the compensated magnetic order, enabling a tunable net magnetization [23,24].

Specially, the piezovalley effect in two-dimensional AM materials is distinguished by its symmetry-driven origin. Unlike conventional valley materials, in which valley pairs are protected by time-reversal symmetry ($T$-paired valleys), the valley degeneracy in AM systems is governed primarily by crystal symmetry, giving rise to so-called $C$-paired valleys [25,26]. This characteristic renders the valley degree of freedom highly susceptible to external perturbations, enabling uniaxial strain to effectively lift valley degeneracy and induce substantial valley polarization. Strain-induced valley polarization has been observed in AM systems such as $Fe_2MoS_4$ and $Ti_2Se_2S$ [27,28]. The anomalous valley Hall effect (AVHE), a direct electrical signature of valley polarization arising from the asymmetric distribution of Berry curvature in momentum space, can be likewise triggered by uniaxial strain. By breaking valley degeneracy, strain enables directional transport of valley-polarized carriers, a mechanism of considerable importance for the development of ultrahigh-speed spintronic and switching devices [29-31].

Although individual piezo-induced effects have been reported in two-dimensional AM materials, their integration within a single material system remains largely unexplored. Furthermore, under conventional valley selection rules, valley polarization in the valence and conduction bands remains coupled, precluding independent control. Strategies to overcome this limitation and achieve independent manipulation of valley polarization, as well as to realize the single-spin-channel AVHE, have yet to be systematically investigated.

In this work, we predict that the Janus monolayer $Cr_2SSe$ serves as a versatile platform for multipiezo effects and valleytronics via first-principles calculations. We identify $Cr_2SSe$ as a semiconductor with an in-plane staggered magnetic ground state, exhibiting large valley splitting and spin-valley locking, wherein the conduction and valence bands at the X and Y valleys are occupied by opposite spin channels. Through systematic symmetry analysis, we show that valley degeneracy at these points is protected by diagonal mirror ($M_{xy}$) symmetry rather than time-reversal symmetry, making the system highly susceptible to uniaxial strain. This symmetry characteristic enables effective control over valley polarization magnitude and orientation, with a maximum polarization of 57.6 meV achieved under 4% tensile strain. Beyond its valley properties, $Cr_2SSe$ exhibits an intrinsic spontaneous electric polarization of 4.13 pC/m, tunable to 4.37 pC/m under external strain, along with a strain induced net magnetization, demonstrating the concurrent activation of piezovalley, piezoelectric, and piezomagnetic responses. Under compressive strains of -2% and -3%, we observe selective reversal of valley polarization and single-spin-channel AVHE. These findings establish a theoretical foundation for the precise manipulation of valley degrees of freedom and provide a pathway for the design of multifunctional valleytronic and spintronic devices.

## II. CALCULATION DETAILS

First-principles calculations were performed within the framework of density functional theory using the Vienna *ab initio* Simulation Package (VASP) [32,33]. The projector augmented wave (PAW) method was employed to describe the electron-ion interaction, and the Perdew-Burke-Ernzerhof (PBE) parametrization of the generalized gradient approximation (GGA) was adopted for the exchange-correlation functional [34]. Structural optimization of the Janus monolayer $Cr_2SSe$ was carried out with convergence criteria of $10^{-6}$ eV for total energy and 0.001 eV Å$^{-1}$ for atomic forces. A plane-wave cutoff energy of 500 eV was adopted, and a vacuum layer of approximately 30 Å was introduced along the out-of-plane direction to eliminate spurious interactions between periodic images [35]. Brillouin-zone integrations were performed using a Gamma-centered Monkhorst-Pack *k*-point grid of 13×13×1 for both structural relaxation and electronic structure calculations. To account for the strong correlation effects of Cr 3*d* electrons, the DFT+*U* method was employed with an effective Hubbard parameter $U_{eff}$ =3.5 eV (where $U_{eff}$ =$U$−$J$, with $J$=0) [36,37]. Thermal stability was assessed through *ab initio molecular dynamics* (AIMD) simulations in a 5×5×1 supercell at 300 K for 5 ps [38]. Dynamic stability was confirmed by phonon dispersion calculations using the PHONOPY code [39]. Maximally localized Wannier functions were constructed using the WANNIER90 package, based on the *d* orbitals of Cr atoms and the *p* orbitals of Se and S atoms [40]. The Berry curvature was subsequently computed from the obtained Wannier interpolation. Electric polarization was evaluated using the Berry-phase method [41,42], and the elastic stiffness tensor $C_{ij}$ was derived within the finite-displacement approach [43].

## III. RESULTS AND DISCUSSION

### A. Stability, Magnetism, and Electronic Structures

Janus monolayer $Cr_2SSe$ crystallizes in a tetragonal structure with space group *p4mm* (No. 99) and with optimized in-plane lattice constants $a = b = 3.92$ Å. The top-view crystal structure reveals a tetragonal lattice formed by Cr, S, and Se atoms, with the unit cell exhibiting $M_{xy}$ symmetry [Fig. 1(a)]. The side view shows the atomic stacking sequence S-Cr-Se along the out-of-plane direction [Fig. 1(b)]. The structural stability of monolayer $Cr_2SSe$ was confirmed through multiple approaches. Phonon dispersion calculations reveal no imaginary frequencies throughout the Brillouin zone, demonstrating dynamic stability [Fig. 1(c)]. AIMD simulations at room temperature show minimal total energy fluctuations over 5 ps, confirming robust thermal stability [Fig. 1(d)]. Furthermore, the calculated elastic constants ($C_{11} = 62.9$ N/m, $C_{12} = 8.2$ N/m) satisfy the Born-Huang criteria ($C_{11} > 0$ and $C_{11} > |C_{12}|$), establishing mechanical stability [44].

To determine the magnetic ground state of Janus monolayer $Cr_2SSe$, we calculated the total energy difference between various FM and AFM configurations with an effective Hubbard parameter $U_{eff} = 3.5$ eV for Cr atoms [21], as shown in Fig. S1 in the Supplemental Material. The AFM1 configuration is energetically most favorable [Fig. S2]. Spin-resolved density of states calculations for this ground state reveal pronounced anisotropy between opposite-spin sublattices [Fig. S3]. Crucially, these sublattices are connected by rotation ($C_{4v}$) and $M_{xy}$ symmetry operations rather than by *PT* symmetry, confirming that Janus monolayer $Cr_2SSe$ hosts an AM order.

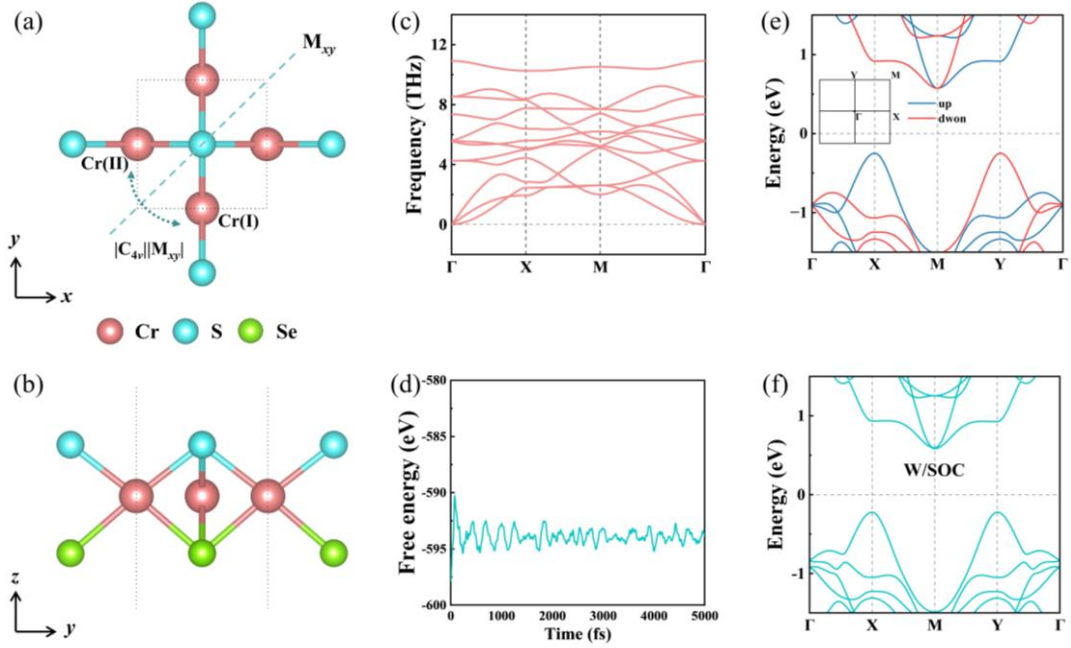

FIG. 1. (a) Top view and (b) side view of the Janus monolayer Cr$_2$SSe structure. Pink, green, and cyan spheres represent Cr, Se, and S atoms, respectively. (c) Phonon dispersion curves. (d) AIMD simulation over 5 ps. (e) Band structure without SOC. (f) Band structure including SOC.

Band structure calculations performed without spin-orbit coupling (SOC) reveal a direct bandgap of 0.82 eV and pronounced staggered spin splitting characteristic of AM order [Fig. 1(e)]. Notably, at the X and Y valleys, the conduction and valence bands are occupied by opposite spin channels, giving rise to intrinsic spin-valley locking. Inclusion of SOC leaves the bandgap largely unchanged [Fig. 1(f)], confirming the non-relativistic origin of the spin splitting, a hallmark of AM [45]. Magnetic anisotropy energy (MAE), defined as MAE=$E_{001}-E_{100}$, was calculated to be 0.10 meV [Fig. S4], indicating an in-plane easy magnetization axis and being consistent with an in-plane staggered magnetic ground state.

**B. Piezovalley effect**

Strain engineering provides a powerful strategy for modulating the physical properties and is particularly effective in two-dimensional AM systems [46]. By reducing lattice symmetry, uniaxial strain can directly tune the coupling between electric, magnetic, and valley degrees of freedom, establishing the material as a multifunctional platform

In Cr$_2$SSe, valley polarization arises from $C$-paired valleys, in contrast to conventional systems where valleys are linked by $T$-paired valleys. This symmetry-driven origin implies that applying uniaxial strain, which breaks the $M_{xy}$ symmetry, can effectively lift valley degeneracy and induce valley polarization. To investigate this, we calculated the band structure of Janus monolayer Cr$_2$SSe under uniaxial strain. Valley polarization is quantified as the energy difference between the X and Y valleys at the conduction-band minimum (CBM) and valence-band maximum (VBM), defined as $\Delta C(V) = E_{XC(V)} - E_{YC(V)}$. Figures 2(a) and 2(b) show the band structure under +2% uniaxial strain applied along the $x$- and $y$-directions, respectively. In both cases, the applied strain breaks $M_{xy}$ symmetry, lifting valley degeneracy and inducing valley polarization. Notably, the strain-induced splitting differs between the conduction and valence bands, leading to distinct valley polarization values for each. Under +2% strain along the $x$-direction, the valence and conduction bands exhibit valley polarizations of -12.4 meV and +28.7 meV, respectively [Fig. 2(c)]. Reversing the strain direction to the $y$-direction flips the sign of the valley polarization, yielding values of +12.4 meV for the valence band and -28.7 meV for the conduction band [Fig. 2(d)]. This sign reversal originates from the $M_{xy}$ symmetry, which dictates opposite evolution trends for valley polarization along the $x$- and $y$-directions.

Moreover, we calculated the band structure under uniaxial strain ranging from −4% to +4% [Figs. S5 and S6]. As shown in Fig. 3(c), the bandgap varies systematically with strain, yet the material retains its AM character across the entire range, demonstrating the robustness of its magnetic ordering under mechanical deformation. The evolution of valley polarization with strain is presented in Figs. 2(e) and 2(f). The results reveal a linear variation of conduction band valley polarization with both compressive and tensile strain, whereas the valence band response is nonlinear. Notably, the trends along the *x*- and *y*-directions are opposite, consistent with the symmetry analysis. Remarkably, under +4% uniaxial strain, we obtain a maximum conduction band valley polarization of 57.6 meV, a value that exceeds those reported for most ferrovalley materials. This large, tunable valley polarization, combined with the ability to control its sign via strain direction, establishes $Cr_2SSe$ as a promising platform for valleytronic applications.

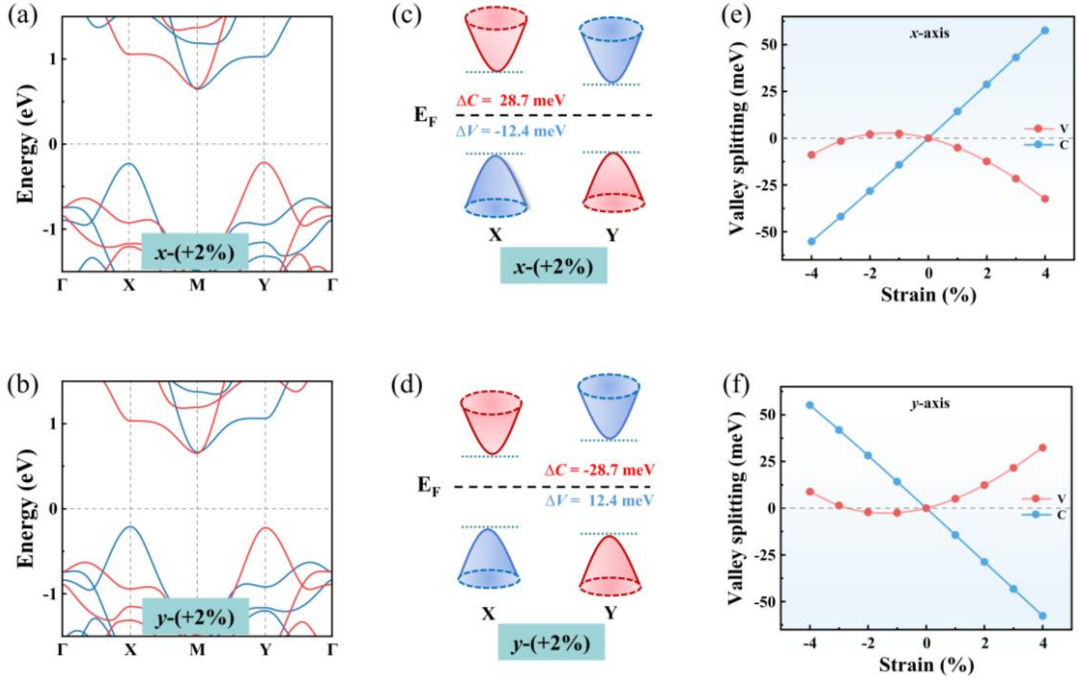

FIG. 2. Band structure and valley polarization of Janus monolayer $Cr_2SSe$ under +2% uniaxial strain applied along the *x*- and *y*-directions. (a, c) Results for strain along the *x*-direction; (b, d) results for strain along the *y*-direction. In the valley polarization diagrams, red and blue cones denote spin-up and spin-down bands, respectively, for the conduction and valence bands. (e, f) Valley polarization as a function of strain along the *x*- and *y*-directions, respectively.

## C. Piezomagnetic effect

In its ground state, the AM order in Janus monolayer Cr$_2$SSe exhibits zero net magnetization, a consequence of protecting symmetries such as $M_{xy}$ symmetry and $C_{4v}$ symmetry. Disrupting these symmetries breaks the spin compensation and can yield a finite net magnetic moment. Uniaxial strain provides a direct means to achieve this, thereby enabling the piezomagnetic effect. When strain is applied, the local chemical environment, spin, and charge distribution around the two inequivalent Cr sublattices, denoted Cr(I) and Cr(II), become distinct. This symmetry lowering induces a differential change in their local magnetic moments, resulting in a non-zero net magnetization. Consequently, the system transitions from a perfectly compensated AM state to a weakly magnetized, FM-like state under applied strain. We calculated the evolution of the net magnetic moment under uniaxial strain ranging from -4% to +4% applied along the *x*- and *y*-directions [Fig. 3(a)]. Both tensile and compressive strains induce a weak net magnetic moment per unit cell, confirming the presence of a piezomagnetic response. To elucidate its microscopic origin, we analyzed the Bader charge distribution on the two Cr sublattices as a function of strain [Fig. 3(b)]. Under -4% strain, the Bader charge on the spin-up Cr(I) atoms increases significantly, by approximately 0.01e, whereas the charge on the spin-down Cr(II) atoms remains largely unchanged. According to Hund's rules [47-49], this strain-induced charge imbalance between the sublattices directly results in a finite net magnetic moment. Notably, this intrinsic piezomagnetic effect is achieved without the need for carrier doping, highlighting the unique susceptibility of the AM order to mechanical perturbations.

## D. Piezoelectric effect

Owing to its non-centrosymmetric structure, lacking both inversion symmetry and interlayer mirror symmetry, Janus monolayer Cr$_2$SSe is expected to exhibit piezoelectric properties [21,22]. To confirm the presence of an intrinsic electric polarization, we first calculated the average electrostatic potential along the out-of-plane direction [Fig. S7]. A pronounced potential difference between the top and bottom surfaces is observed, indicative of a built-in polarization field. Quantifying this using the Berry-phase method yields a spontaneous electric polarization of 4.13 pC/m. We next investigated the strain dependence of this electric polarization. As shown in Fig. 3(d), applying uniaxial strain along the *x*-direction over a range of −4% to +4% reveals a linear relationship between electric polarization and strain, a hallmark of the piezoelectric effect. Notably, under -4% compressive strain, the electric polarization increases to 4.37 pC/m. This value is substantially higher than those reported for many two-dimensional piezoelectric materials [50], underscoring the strong electromechanical coupling in Janus monolayer Cr$_2$SSe.

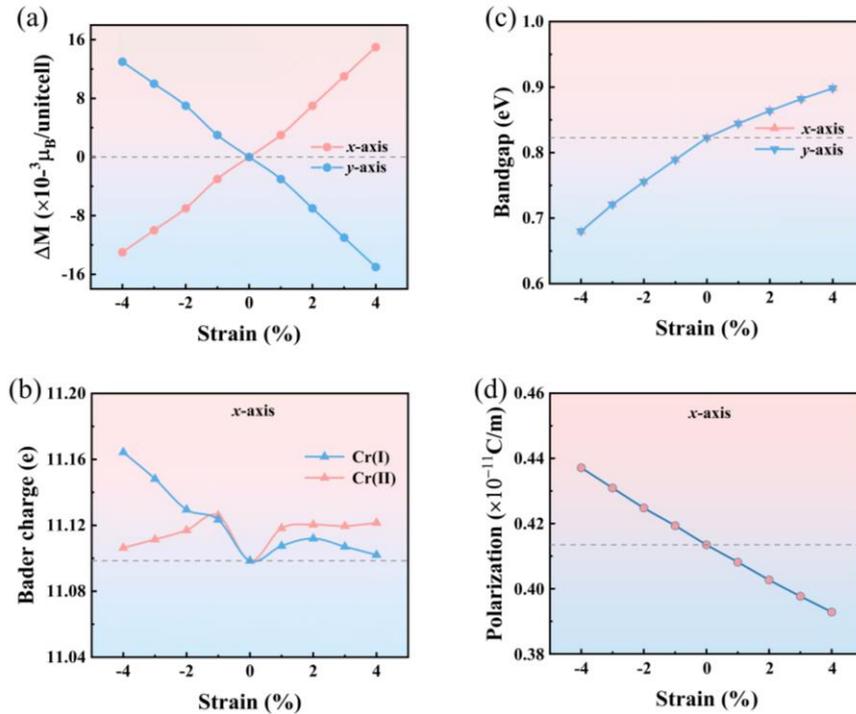

FIG. 3. (a) Net magnetic moment of Janus monolayer Cr$_2$SSe as a function of uniaxial strain. (b) Bader charge variation on Cr atoms under uniaxial strain. (c) Bandgap evolution under uniaxial strain. (d) Electric polarization as a function of uniaxial strain.

**E. Selective Reversal of Valley Polarization**

As noted earlier, the valley polarization of Janus monolayer Cr$_2$SSe exhibits a nonlinear response at the valence band under applied strain. To elucidate this behavior, we investigated the effects of −2% and −3% uniaxial strain applied along the $x$-direction [Figs. 4(a) and 4(b)]. At -2% strain, the valley polarization values are -28.2 meV for the conduction band and +2.2 meV for the valence band [Fig. 4(c)]. Increasing the strain to -3% shifts these values to -41.9 meV and -1.6 meV, respectively [Fig. 4(d)]. Notably, while the magnitude of conduction band valley polarization increases monotonically with strain, its sign remains unchanged. By contrast, the valence band valley polarization undergoes a sign reversal between -2% and -3% strain. This decoupled response demonstrates that uniaxial strain can selectively reverse the valley polarization of a single valley without affecting the valley polarization direction of the other, enabling independent control over the valley degrees of freedom in different bands.

Furthermore, we elucidated the mechanism underlying the decoupled valley polarization response by calculating the orbital-projected band contributions under −2% and −3% strain [Fig. S8]. The valence band is predominantly composed of Se $p$ orbitals, whereas the conduction band arises primarily from Cr $d$ orbitals. This difference in orbital character underpins their distinct strain sensitivities, enabling selective reversal of valley polarization [51,52]. According to perturbation theory, $p$ orbitals possess a dumbbell-shaped geometry with directional electron density distributed along specific axes, rendering them highly sensitive to lattice strain, particularly when the strain is applied along the orbital axis. Such deformation alters orbital overlap integrals and modifies the band structure. In contrast, $d$ orbitals exhibit more complex, multi-directional electron distributions (e.g., flower-like patterns), resulting in an intermediate strain sensitivity that depends on both the magnitude and direction of the applied strain. This disparity in orbital response accounts for the observed independent tunability of the valence and conduction band valleys.

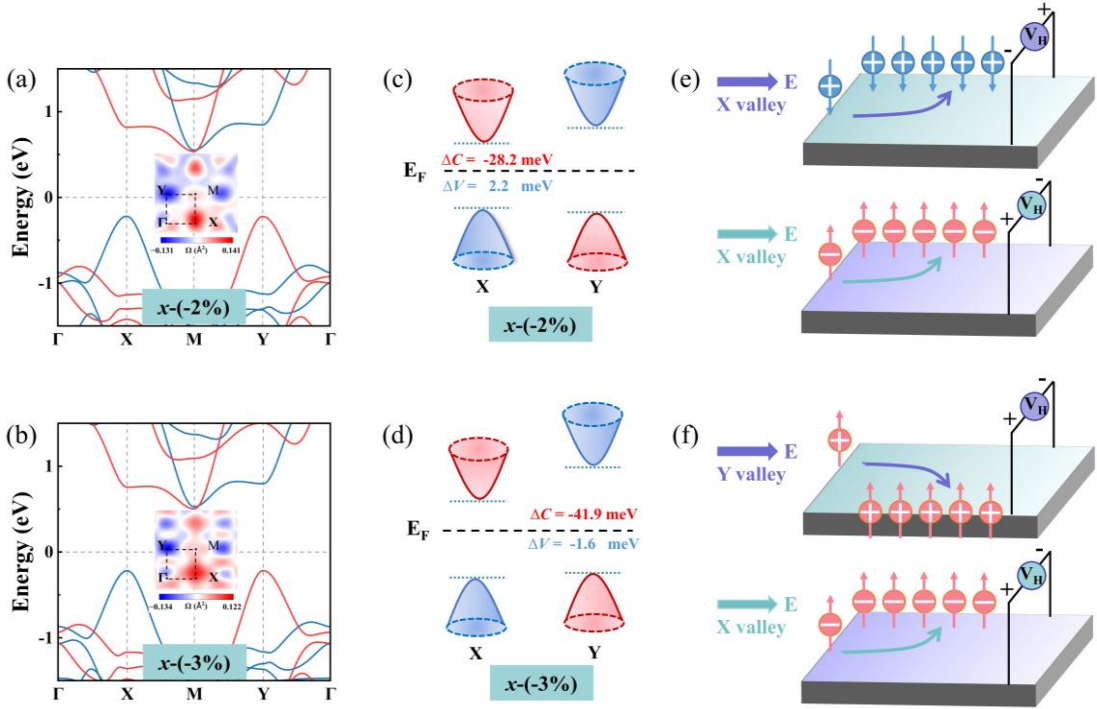

FIG. 4. Band structure, Berry curvature, valley polarization, and AVHE of Janus monolayer Cr$_2$SSe under uniaxial strain of -2% and -3% applied along the *x*-direction. (a, c, e) Results for -2% uniaxial strain; (b, d, f) results for -3% uniaxial strain. In the AVHE schematics, arrows indicate spin orientation, with holes and electrons represented by "+" and "−", respectively.

**F. Single-Spin-Channel Anomalous Valley Hall Effect**

In two-dimensional systems, the Berry curvature acts as an effective pseudomagnetic field, enabling the AVHE to manifest without an external magnetic field. To characterize the valley-contrasting and spin transport properties of Janus monolayer Cr$_2$SSe, we calculated its Berry curvature using the Kubo formula [53]:

$$\Omega_z(k) = -\sum_n \sum_{n \neq m} f_n(\bar{k}) \frac{2 \operatorname{Im} \langle \psi_{nk} | \hat{v}_x | \psi_{mk} \rangle \langle \psi_{mk} | \hat{v}_y | \psi_{nk} \rangle}{(E_{nk} - E_{mk})^2} \qquad (1)$$

where $f_n(k)$ is the Fermi-Dirac distribution function, $\psi_{nk}$ ($\psi_{mk}$) the Bloch wave function with eigenvalue $E_{nk}$ ($E_{mk}$), $\hat{v}_{x/y}$ is the velocity operator along the $x/y$ direction, respectively. In its pristine state, Janus monolayer Cr$_2$SSe preserves $M_{xy}$ symmetry, resulting in a net Berry curvature that vanishes throughout the Brillouin zone. Application of uniaxial strain breaks this symmetry, inducing a finite Berry curvature. The insets in Figs. 4(a) and 4(b) illustrate the Berry curvature distribution under -2% and -3% uniaxial strain applied along the $x$-direction. At -2% strain, the Berry curvature at the X and Y valleys is approximately 0.141 Å$^2$ and -0.131 Å$^2$, respectively. At -3% strain, following the inversion of the valence band valley, and the Berry curvature becomes of 0.122 Å$^2$ at X valley and -0.134 Å$^2$ at Y valley. These results demonstrate that uniaxial strain effectively controls both the magnitude and sign of the Berry curvature, providing a route to manipulate the AVHE. The Berry curvature exhibits opposite signs and unequal magnitudes at the X and Y valleys. In the presence of an electric field, this Berry curvature imparts an anomalous velocity to charge carriers, given by $v \approx -\frac{e}{\hbar} E \times \Omega(k)$ [29]. This anomalous velocity drives directional carrier deflection, leading to charge accumulation at the sample boundaries and the generation of a Hall voltage.

As illustrated in Fig. 4(e), under -2% strain applied along the $x$-direction and under hole (or electron) doping, the Fermi level shifts toward the valence (or conduction) band. The Berry curvature imparts an anomalous velocity such that spin-down holes in the X valley (or spin-up electrons) acquire a lateral velocity and accumulate at one edge, generating a Hall voltage and inducing the AVHE. In stark contrast, when the applied $x$-direction strain is increased to -3%, strain driven selective valley inversion fundamentally remodels the valley- and spin-resolved transport behavior. This valley inversion dictates that both Y valley holes and X valley electrons are transported exclusively through a single spin-up channel, with the opposite spin-down channel making a negligible contribution to carrier transport regardless of doping type. This unique valley-spin locked transport regime directly gives rise to a single-spin-channel AVHE, defined as the regime where all charge carriers participating in conduction possess a 100% uniform spin orientation (here, all spin-up), with carriers of the opposite spin contributing nothing [Fig. 4(f)]. This strain-tunable, single-spin-channel AVHE thus bridges fundamental quantum transport physics with next-generation valleytronic and spintronic devices.

## IV. CONCLUSION

In summary, we identify Janus monolayer $Cr_2SSe$ as a stable two-dimensional AM material that intrinsically combines spin-valley locking with spontaneous electric polarization. The material exhibits piezovalley, piezoelectric, and piezomagnetic responses that are concurrently tunable under uniaxial strain. By breaking the in-plane $M_{xy}$ symmetry, deterministic control over valley degrees of freedom is achieved: under +4% strain, the valley polarization reaches 57.6 meV, and reversing the strain direction flips the valley polarization sign. More significantly, we observe selective reversal of valley polarization: under compressive strains of -2% and -3%, the valence band valley polarization reverses while the conduction band polarization remains unchanged, enabling decoupled and independent control of valley degrees of freedom across bands. The strain induced Berry curvature also gives rise to a single-spin-channel AVHE under -3% strain. These findings establish Janus monolayer $Cr_2SSe$ as a unique platform for synergistic multipiezo effects, providing a foundation for low-power valleytronics and quantum sensing, and offering new strategies for multifunctional valley control in two-dimensional systems.

**ACKNOWLEDGMENTS**

This work was supported by the National Natural Science Foundation of China (Grant Nos. 12364007, 12264016) and the Scientific Research Foundation of Hunan Provincial Education Department, China (Grant No. 24A0366). And the computing work is supported by the Open Source Supercomputing Center of S-A-I.

**SUPPLEMENTARY MATERIAL**

See the supplementary material for the magnetic configurations, MAE, band structures, and effective potential profiles, etc.

**References**

[1] X. Chen, Y. Liu, P. Liu, Y. Yu, J. Ren, J. Li, A. Zhang, and Q. Liu, Unconventional magnons in collinear magnets dictated by spin space groups, Nature **640**, 349 (2025).

[2] X. Chen, J. Ren, Y. Zhu, Y. Yu, A. Zhang, P. Liu, J. Li, Y. Liu, C. Li, and Q. Liu, Enumeration and representation theory of spin space groups, Phys. Rev. X **14**, 31038 (2024).

[3] Y. Jiang, Z. Song, T. Zhu, Z. Fang, H. Weng, Z. X. Liu, J. Yang, and C. Fang, Enumeration of spin-space groups: Toward a complete description of symmetries of magnetic orders, Phys. Rev. X **14**, 31039 (2024).

[4] M. Naka, S. Hayami, H. Kusunose, Y. Yanagi, Y. Motome, and H. Seo, Spin current generation in organic antiferromagnets, Nat. Commun. **10**, 4305 (2019).

[5] H. Y. Ma, M. Hu, N. Li, J. Liu, W. Yao, J. F. Jia, and J. Liu, Multifunctional antiferromagnetic materials with giant piezomagnetism and noncollinear spin current, Nat. Commun. **12**, 2846 (2021).

[6] L. Šmejkal, J. Sinova, and T. Jungwirth, Beyond conventional ferromagnetism and antiferromagnetism: A phase with nonrel-ativistic spin and crystal rotation symmetry, Phys. Rev. X **12**,031042 (2022).

[7] I. Mazin, Altermagnetism-a new punch line of fundamental magnetism, Phys. Rev. X **12**, 040002 (2022).

[8] H. Sun, P. Dong, C. Wu, and P. Li, Multifield-induced antiferromagnet transformation into altermagnet and realized anomalous valley Hall effect in monolayer VPSe$_3$, Phys. Rev. B **111**, 235431 (2025).

[9] H. Chen, F. Chen, H. Cheng, X. Zhao, G. Hu, X. Yuan, and J. Ren, Layer-locked multiple valley Hall effects in tetragonal altermagnetic/ferromagnetic monolayers M$_2$SiCX$_2$ (M = transition metal ; X = S , Se ), Phys. Rev. B **111**, 155428 (2025).

[10] Y. Jiang, X. Zhang, H. Bai, Y. Tian, B. Zhang, W. J. Gong, and X. Kong, Strain-engineering spin-valley locking effect in altermagnetic monolayer with multipiezo properties, Appl. Phys. Lett. **126**, 053102 (2025).

[11] W. Xie, X. Xu, Y. Yue, H. Xia, and H. Wang, Piezovalley effect and magnetovalley coupling in altermagnetic semiconductors studied by first-principles calculations, Phys. Rev. B **111**, 134429 (2025).

[12] Y. Zhang, Y. Liu, and Z. L. Wang, Fundamental theory of piezotronics, Adv. Mater. **23**, 3004 (2011).

[13] X. Wen, W. Wu, C. Pan, Y. Hu, Q. Yang, and Z. L. Wang, Development and progress in piezotronics, Nano Energy **14**, 276 (2015).

[14] J. Jiang, S. Liu, L. Feng, and D. Zhao, A review of piezoelectric vibration energy harvesting with magnetic coupling based on different structural characteristics, Micromachines **12**, 436 (2021).

[15] C. W. Nan, Magnetoelectric effect in composites of piezoelectric and piezomagnetic phases, Phys. Rev. B **50**, 6082 (1994).

[16] D. Davino, A. Giustiniani, and C. Visone, The piezo-magnetic parameters of terfenol-d: An experimental viewpoint, Phys. B: Condens. Matter **407**, 1427 (2012).

[17] M. Ikhlas, S. Dasgupta, F. Theuss, T. Higo, S. Kittaka, B. J. Ramshaw, O. Tchernyshyov, C. W. Hicks, and S. Nakatsuji, Piezomagnetic switching of the anomalous Hall effect in an antiferromagnet at room temperature, Nat. Phys. **18**, 1086 (2022).

[18] K. F. Mak, K. L. McGill, J. Park, and P. L. McEuen, The valley Hall effect in MoS$_2$ transistors, Science **344**, 1489 (2014).

[19] D. Xiao, G. B. Liu, W. X. Feng, X. D. Xu, and W. Yao, Coupled spin and valley

physics in monolayers of MoS$_2$ and other group VI dichalcogenides, Phys. Rev. Lett. **108**, 196802 (2012).

[20] X. D. Xu, W. Yao, D. Xiao, and T. F. Heinz, Spin and pseudospins in layered transition metal dichalcogenides, Nat. Phys. **10**, 343 (2014).

[21] I. Khan, D. Bezzerga, B. Marfoua, and J. Hong, Altermagnetism, piezovalley, and ferroelectricity in two-dimensional Cr$_2$SeO altermagnet, npj 2D Mater. Appl. **9**, 18 (2025).

[22] S. D. Guo, X.-S. Guo, K. Cheng, K. Wang, and Y. S. Ang, Piezoelectric altermagnetism and spin-valley polarization in Janus monolayer Cr$_2$SO, Appl. Phys. Lett. **123**, 082401 (2023).

[23] X. Xu and L. Yang, Alterpiezoresponse in two-dimensional Lieb-Lattice altermagnets, Nano Lett. **25**, 11870 (2025).

[24] W. Zhang, B. Xiao, C. Li, C. Qiu, H. Zeng, and J. Zhao, Multiple strain-induced effects beyond the piezoelectric effect in altermagnetic monolayer Co$_2$MoSe$_4$, Phys. Rev. B **112**, 144436 (2025).

[25] H. Y. Ma, M. Hu, N. Li, J. P. Liu, W. Yao, J. F. Jia, and J.W. Liu, Multifunctional antiferromagnetic materials with giant piezomagnetism and noncollinear spin current, Nat. Commun. **12**, 2846 (2021).

[26] M. Naka, Y. Motome, and H. Seo, Perovskite as a spin current generator, Phys. Rev. B **103**, 125114 (2021).

[27] Q. Shen, W. Liao, H. Bao, D. Xu, J. Tan, J. Dong, and G. Ouyang, Tunable valley polarization and spin splitting in altermagnetic Fe$_2$MoS$_2$X$_2$ (X = S , Se , Te) via symmetry engineering, Phys. Rev. Mater. **9**, 114404 (2025).

[28] X. Hu, W. Zhao, W. Xia, H. Sun, C. Wu, Y.-Z. Wu, and P. Li, Valley polarization and anomalous valley Hall effect in altermagnet Ti$_2$Se$_2$S with multipiezo properties, Appl. Phys. Lett. **127**, 011905 (2025).

[29] D. Xiao, M. C. Chang, and Q. Niu, Berry phase effects on electronic properties, Rev. Mod. Phys. **82**, 1959 (2010).

[30] H. J. Kim, C. Li, J. Feng, J. H. Cho, and Z. Zhang, Competing magnetic orderings and tunable topological states in two-dimensional hexagonal organometallic

lattices, Phys. Rev. B **93**, 41404 (2016).

[31] A. Kormányos, V. Zólyomi, V. I. Fal'ko, and G. Burkard, Tunable Berry curvature and valley and spin Hall effect in bilayer MoS$_2$, Phys. Rev. B **98**, 35408 (2018).

[32] G. Kresse and J. Hafner, *Ab initio* molecular dynamics for liquid metals, Phys. Rev. B **47**, 558 (1993).

[33] G. Kresse and J. Hafner, *Ab initio* molecular-dynamics simulation of the liquid-metal–amorphous-semiconductor transition in germanium, Phys. Rev. B **49**, 14251 (1994).

[34] J. P. Perdew, K. Burke, and M. Ernzerhof, Generalized gradient approximation made simple, Phys. Rev. Lett. **77**, 3865 (1996).

[35] S. Grimme, S. Ehrlich, and L. Goerigk, Effect of the damping function in dispersion corrected density functional theory, J. Comput. Chem. **32**, 1456 (2011).

[36] V. I. Anisimov, J. Zaanen, and O. K. Andersen, Band theory and mott insulators: hubbard *U* instead of stoner *I*, Phys. Rev. B **44**, 943 (1991).

[37] A. I. Liechtenstein, V. I. Anisimov, and J. Zaanen, Density-functional theory and strong interactions: Orbital ordering in Mott-Hubbard insulators, Phys. Rev. B **52**, R5467 (1995).

[38] S. Nosé, A unified formulation of the constant temperature molecular dynamics methods, J. Chem. Phys. **81**, 511 (1984).

[39] D. Alfè, PHON: A program to calculate phonons using the small displacement method, Comput. Phys. Commun. **180**, 2622 (2009).

[40] A. A. Mostofi, J. R. Yates, Y.-S. Lee, I. Souza, D. Vanderbilt, and N. Marzari, Wannier90: A tool for obtaining maximally-localised wannier functions, Comput. Phys. Commun. **178**, 685 (2008).

[41] R. D. King-Smith and D. Vanderbilt, Theory of polarization of crystalline solids, Phys. Rev. B **47**, 1651 (1993).

[42] R. Resta, M. Posternak, and A. Baldereschi, Towards a quantum theory of polarization in ferroelectrics: The case of KNbO$_3$, Phys. Rev. Lett. **70**, 1010 (1993).

[43] X. Wu, D. Vanderbilt, and D. R. Hamann, Systematic treatment of

displacements, strains, and electric fields in densityfunctional perturbation theory, Phys. Rev. B **72**, 035105 (2005).

[44] R. C. Andrew, R. E. Mapasha, A. M. Ukpong, and N. Chetty, Mechanical properties of graphene and boronitrene, Phys. Rev. B **85**, 125428 (2012).

[45] Y. Q. Li, Y. K. Zhang, X. L. Lu, Y. P. Shao, Z. Q. Bao, J. D. Zheng, W. Y. Tong, and C. G. Duan, Ferrovalley physics in stacked bilayer altermagnetic systems, Nano Lett. **25**, 6032 (2025).

[46] W. Xun, X. Liu, Y. Zhang, Y. Z. Wu, and P. Li, Stacking-, strain-engineering induced altermagnetism, multipiezo effect, and topological state in two-dimensional materials, Appl. Phys. Lett. **126**, 161903 (2025).

[47] J. Katriel and R. Pauncz, Theoretical interpretation of Hund's Rule, Adv. Quantum Chem. **10**, 143 (1977).

[48] J. W. Warner and R. S. Berry, Hund's rule, Nature **313**, 160 (1985).

[49] W. Kutzelnigg and J. D. Morgan, Hund's rules, Z. Phys. D: At., Mol. Clusters **36**, 197 (1996).

[50] Y. Cui, T. Wang, D. Hu, Z. Wang, J. Hong, and X. Wang, Piezoelectricity in $NbOI_2$ for piezotronics and nanogenerators, npj 2D Mater. Appl. **8**, 62 (2024).

[51] C. G. Van De Walle, Band lineups and deformation potentials in the model-solid theory, Phys. Rev. B **39**, 1871 (1989).

[52] J. Bardeen and W. Shockley, Deformation potentials and mobilities in non-polar Crystals, Phys. Rev. **80**, 72 (1950).

[53] D. J. Thouless, M. Kohmoto, M. P. Nightingale, and M. Den Nijs, Quantized Hall conductance in a two-dimensional periodic potential, Phys. Rev. Lett. **49**, 405 (1982).